\documentstyle[12pt,psfig,epsfig]{article}
\newcommand{\beq}{\begin{equation}}
\newcommand{\eeq}{\end{equation}}
\newcommand{\beqarray}{\begin{eqnarray}}
\newcommand{\eeqarray}{\end{eqnarray}}
\def\lsim{\raise0.3ex\hbox{$\;<$\kern-0.75em\raise-1.1ex\hbox{$\sim\;$}}}
\def\gsim{\raise0.3ex\hbox{$\;>$\kern-0.75em\raise-1.1ex\hbox{$\sim\;$}}}
\def\para{\vspace{0.3cm}\noindent}

\begin{document}
\begin{center}
{\large \bf The  Appearance of Tau Neutrinos from a Gamma Ray Burst  
  }

\medskip

{Nayantara Gupta \footnote{tpng@mahendra.iacs.res.in}}\\  
{\it Department of Theoretical Physics,\\
 Indian Association for the Cultivation of Science,\\
Jadavpur, Kolkata 700 032, INDIA.}  

\end{center}

\begin{abstract}
The muon neutrinos from the Gamma Ray Bursts (GRBs) during their propagation oscillate to tau neutrinos.  The tau neutrinos can be identified by the appearance
of tau leptons, which decay to muons below the detector.
 We examine the prospect of detecting these tau neutrinos from individual GRBs in a large area muon detector.
We investigate for what range of values of the physical parameters of the GRBs
such as Lorentz factor, redshift (distance from the observer) and total energy
released in neutrino emissions the appearance of tau neutrinos from individual
GRBs will be observable in a ground based muon detector of $km^{2}$ area.
We find that it will be possible to detect tau neutrino signals from a single GRB in a ground based muon detector of $km^{2}$ area above a muon threshold energy 30 PeV and at a zenith angle of $\theta=180^{\circ}$ if the energy emitted in neutrinos by the GRB is $10^{53}erg$ and its redshift is below 0.01 or if the energy emitted in neutrinos by the GRB is $10^{54}erg$ and its redshift is below 0.03.
\end{abstract}

PACS number(s): 95.55.Vj
\newpage
\section{Introduction}
In proton acceleration models the intrinsically produced tau neutrino flux is
expected to be very small, typically a factor between $10^{-5}$ and $10^{-6}$ relative to electron and muon neutrino fluxes. The recent experimental measurements of atmospheric neutrinos suggest that neutrinos could just have vacuum flavor
oscillations and the tau neutrino flux would be enhanced. 
In atmospheric neutrinos the contribution of tau neutrinos would be small but it would not be so in case of extragalactic neutrinos.
The appearance of high energy tau neutrinos due to $\nu_{\mu}\rightarrow \nu_{\tau}$ oscillations can be observed by measuring the neutrino induced upward hadronic and electromagnetic showers or upward going muons. 

\para
The muon and electron neutrinos during their propagation through the Earth are absorbed due to charged current and neutral current interactions.
We can expect an exponential fall of the numbers of muon and electron neutrinos  at high energies due to absorptions after their passage through the Earth. The high energy tau neutrinos are also absorbed in the matter of the Earth but they are regenerated.
The tau leptons produced from $\nu_{\tau}$s in charged current interactions
again decay to produce $\nu_{\tau}$s of lower energies. 
The $\tau$ leptons produced from $\nu_{\tau}$s in charged current interactions
decay in several channels. There are six decay modes of $\tau$ leptons described in \cite{sharada1}.In all these decay modes $\nu_{\tau}$s are produced but in
only one of these decay modes muons are produced which are expected to be detected by large area muon detectors. 

\para
It is possible to detect high energy tau neutrinos by their characteristic double shower events \cite{athar} or by the pileup of tau neutrino events near 100 TeV \cite{hal1}.
It has been pointed out in \cite{hal1} the observation of double-bang events is difficult in a first generation telescope such as AMANDA \cite{amanda}. 
 High energy tau neutrinos initiate a cascade in the Earth which will generate tau neutrinos of reduced energies in each interaction. Once the energies of the tau neutrinos fall below the threshold for absorption they propagate directly to the detector with  energies in the range of 10 TeV to 100 TeV.
If the source neutrino energy spectrum falls slower than $E^{-2}$
all $\nu_{\tau}$s with energies greater than 100 TeV will have their energies reduced to about 100 TeV after their passage through the Earth. 
In this way a pileup of tau neutrino events is expected to be produced near neutrino energy  100 TeV.

\para 
  The  very high energy $\nu_{\mu}$ neutrinos are absorbed as they penetrate the Earth.
The $\nu_{\mu}$ neutrinos which pass through a lesser amount of matter inside
the Earth are absorbed in lesser amount compared to the neutrinos which directly cross the diameter of the Earth. The muon neutrino  events are maximum in the horizontal direction and minimum in the vertical direction in the muon detector. These events gradually increase from the vertical direction to the horizontal direction. But we expect the tau neutrino events to be uniformly distributed over all zenith angles in the muon detector because they are regenerated during their propagation through the Earth \cite{hal1}. This is another experimental signature for $\nu_{\tau}$ neutrinos.

\para
In this work
we calculate the upward going muon event rates from muon and tau neutrinos of a GRB for very high muon threshold energies of the detector when the muon neutrinos are mostly absorbed during their propagation through the Earth. We can not confirm the observation of appearance of tau neutrinos in the muon detector unless we reach the threshold energy of the detector when the Earth is no more transparent to muon neutrinos. Moreover the muon neutrinos are not significantly absorbed in the Earth if they are coming from horizontal directions.
In this case the observation of muons even at very high energies does not provide conclusive evidence of the appearance of tau neutrinos from a single GRB.  
So the zenith angle at which the GRB is observed also plays a crucial role in providing unmistakeable signature of the tau neutrinos in the muon detector from a single GRB. The proposed kilometer scale IceCube neutrino telescope from its South Pole location will observe a particular GRB at a fixed zenith angle. In that case we do not have the opportunity to observe uniform tau neutrino event rates at all zenith angles.
We show that if we receive upward going muons at a very high threshold energy above 30PeV from a GRB then only we can confirm the appearance of tau neutrinos from that GRB.
At such high threshold energies the role of atmospheric noise is negligible.
We will be able to conclude about the appearance of tau neutrinos from a GRB
if we can separate the $\nu_{\tau}$ signals from $\nu_{\mu}$ signals of that GRB
 using the phenomena of $\nu_{\mu}$ absorptions during the propagation of $\nu_{\mu}$s through the Earth.
Also if the burst is bright and nearby then only we expect to detect muon signals from that burst.

\para
We have not considered the tau neutrinos which
produce double bang events in this work because they are much less compared to
the events in which muons are produced via tau lepton decays $(\nu_{\tau}\rightarrow \tau \rightarrow {\mu})$ \cite{alv1}.
   
\para
A new effect has been pointed out in \cite{beacom}. The ${\nu_{\tau}}\rightarrow {\tau} \rightarrow {\nu_{\tau}}$ regeneration process creates a secondary
$\bar\nu_{\mu}$ flux. Though the $\bar\nu_{\mu}$ flux is at most 0.2 of the $\nu_{\tau}$ flux, the neutral current channels in total ($\bar\nu_{\mu}$ and $\bar\nu_{e}$ channels) increase the detectability of $\nu_{\tau}$ events by $40\%$.

\para
We proceed as follows in this paper.
In section 2 we mention about the procedure we have used for calculating the number of tau neutrinos from a GRB. 
The formulation used in this work to calculate the number of secondary muon events from tau neutrinos of a GRB has been discussed in section 3. The results
of this work have been explained in section 4.
We find it is possible to detect unmistakeably the signature of tau neutrinos from individual GRBs at zenith angle $180^{\circ}$
in a muon detector of $km^{2}$ area if the total energy emitted by the GRB in neutrino emissions ($E_{GRB}$) is $10^{53}erg$ and its redshift ($z$) is below 0.01 or if $E_{GRB}=10^{54}erg$ and the redshift $z$ of the GRB is below 0.03. 
We need a muon threshold energy 30 PeV for detecting tau neutrino signals in a muon detector of $km^{2}$ area.
One can obtain conditions for detectability of tau neutrino events from a single GRB in $km^{2}$ area muon detector for other values of $E_{GRB}$.
\section{The Number of Tau Neutrinos Produced from Muon Neutrinos of a Gamma Ray Burst}

We consider the fireball model of GRBs (Gamma Ray Bursts) \cite{Eli} and
calculate the expected tau neutrino event rates in a large area muon detector
for different choices of the GRB parameters. Each GRB is characterised by its
Lorentz boost factor $\Gamma$, total energy emitted in neutrinos $E_{GRB}$,
distance of the GRB from the observer (redshift) $z$ and duration of the burst. Also there are other
parameters like photon spectral break energy $E_{\gamma, MeV}^{b}$ and wind variability time $t_{v}$. There is correlation between the parameters minimum Lorentz factor $\Gamma$, wind variability time $t_{v}$ and $E_{{\gamma}, MeV}^{b}$ for  known values of wind luminosity and equipartition parameter $\epsilon_{B}$ \cite{Guetta}. We have used this correlation in calculating the number of neutrinos emitted from a GRB.
The detail procedure for calculation of number of muon neutrinos produced from a GRB has been discussed in \cite{nayan}.
The original muon neutrinos from GRBs are undergoing oscillations during
propagation and tau neutrinos are produced in this way. 
The probablity of oscillation from $\nu_{\mu}$ to $\nu_{\tau}$ is given by
\beq
Prob(\nu_{\mu}\rightarrow\nu_{\tau})=sin^{2}2{\theta}sin^{2}(\frac{{\triangle}m^{2} d}
{4E_{\nu}})
\eeq

Considering the values of neutrino mass difference and mixing as found from SuperKamiokande  data  $(sin^{2}2{\theta}\sim 1, {\triangle}m^{2}\sim 10^{-3} eV^{2})$  \cite{sup} and the distance ($d$) of the source of the order of a megaparsec to thousands of megaparsecs   
the above expression for probablity averages to about half for all relevant neutrino energies to be considered for detection \cite{sharada1}.  
Independent of the energies of the high-energy muon neutrinos their number is suppressed by a costant factor of half due to neutrino oscillations.
The number of muon neutrinos produced from a GRB when multiplied by the probablity of oscillation from muon neutrinos to tau neutrinos gives the number of tau
neutrinos produced from that GRB. 

Number of $\nu_{\tau}$s produced from a GRB due to $\nu_{\mu}\rightarrow \nu_{\tau}$  oscillation = Number of $\nu_{\mu}$s  from the GRB $\times  Prob(\nu_{\mu}\rightarrow \nu_{\tau})$

\section{The Number of Secondary Muons Produced from Tau Neutrinos and Muon Neutrinos of a GRB} 
The tau neutrinos produce tau leptons in charged current interactions during their passage through the Earth. The energy dependent charged current cross-sections for productions of tau leptons from tau neutrinos and muons from muon neutrinos have been used from \cite{raj}. The energy dependent total cross-sections for muon neutrino absorptions have also been used from \cite{raj}.
We have calculated the number of secondary muons from a single GRB produced via the decays of the tau leptons from $\nu_{\tau}$s  using the formula for muon event rate calculation discussed in \cite{sharada1}.
The number of muon events has been calculated as a function of the muon threshold energy of the detector of $km^{2}$ area.
The probablity of the decay process $\tau \rightarrow \nu_{\tau} \mu \bar\nu_{\mu}$ in which a muon will be produced is 0.18 \cite{sharada1}.
The tau lepton carry a fraction 0.75 of the initial tau neutrino energy and the muon produced from $\tau$ decay carry a fraction 0.4 of the tau lepton energy \cite{gaisser, sharada1, beacom}. 
The secondary muons produced from $\tau$ decays below the detector lose energy
during their propagation. We have used the formula for energy loss of muons in
rock from \cite{dar} in calculating the energy dependent ranges of the muons in
rock. 
The procedure for calculation of number of secondary muon events produced from muon neutrinos of a GRB has been discussed in \cite{nayan}.

\section{Results and Discussions}

We study the appearance of tau neutrinos from a single GRB for the first time. Each gamma ray burst is an individual phenomenon of short duration in the Universe. The observation of the tau leptons from GRBs can be used to probe studies on neutrino oscillations with $\triangle m^{2}$ to a value as low as $10^{-17}eV^{2}$ \cite{hal1}. Here the square of the mass difference is relative to the
original $\nu_{e,\mu}$.
This kind of an observation on $\nu_{\tau}$ appearance would extend the present
understanding of neutrino oscillations ($\nu_{\mu}\rightarrow \nu_{\tau}$) by fourteen orders of magnitude.
Neutrino astronomy with tau neutrinos also opens an opportunity to know about
the ultra high energy neutrino productions from different astrophysical objects.As has been discussed in \cite{hal1} if there exist large angle vacuum neutrino
oscillations involving $\nu_{\tau}$ with  sufficient value of $\triangle m^{2}$ consistent with Super-Kamiokande results \cite{super} a pure $\nu_{\mu}$ source of energy upto even $10^{21} eV$ can be observed in the muon detector. Since these neutrinos are of very high energy the final $\nu_{\tau}$s always point back to their source.

\para

In Fig.1. the numbers of muon events from muon neutrinos and tau neutrinos of a GRB have been plotted for different muon threshold energies of the muon detector
of $km^{2}$ area.
The energy of the GRB is assumed to be $10^{54} erg$ and the burst is observed at zenith angle $\theta=180^{\circ}$.The GRB is at a distance of redshift  $z=0.03$. The wind luminosity is assumed to be $10^{53} erg/sec$.
Nearly $10\%$ of the initial fireball proton energy or wind energy is released
in neutrino emissions from a GRB \cite{nayan}. 
The Lorentz factor of the GRB is assumed to be 199.52 for photon spectral break
 energy $E_{\gamma}^{b}=50.118MeV$ \cite{Guetta}.
 We are interested to know the value of muon threshold energy of the detector at which the number of muon events from tau neutrinos of a GRB outnumbers the number of muon events from muon neutrinos of that GRB.  
We have varied the threshold energy of the muon detector upto 100 PeV. As we increase the muon threshold energy above 20PeV we observe the number of muon events from tau neutrinos remains constant at a value of nearly 1event/$km^{2}$.  
 Near muon threshold energy 30PeV the number of muon events from tau neutrinos
of a GRB is more than the number of muon events from muon neutrinos of that GRB.
Also the signal from tau neutrinos does not drop below 1event/$km^{2}$. 
In Fig.1. we have also plotted the sums of the numbers of muon neutrino and tau neutrino events from the GRB in a muon detector of $km^{2}$ area against the muon threshold energy of the detector.   

\para

In Fig.2. the numbers of muon events from muon neutrinos and tau neutrinos of a GRB in a muon detector of $km^{2}$ area have been again plotted as a function of muon threshold energy of the detector. The GRB is assumed to be at a distance of redshift $z=0.01$ and total energy emitted in neutrinos is $E_{GRB}=10^{53}erg$. 
It has been shown in \cite{frail} that the mean isotropic equivalent energy emitted by GRBs is of the order of $10^{53}erg$.
 The allowed values of Lorentz factor and corresponding photon spectral break energy have been used from \cite{Guetta}. 
An initial tau neutrino threshold energy of 100 TeV corresponds to a secondary muon threshold energy of 30 TeV because the muons produced via decays of tau leptons produced from tau neutrinos carry about $3/10$ of the initial tau neutrino energy \cite{beacom}.
In this plot we find that near muon threshold energy 30PeV the number of muon events from tau neutrinos of a GRB remains near 1event/$km^{2}$ while the number muon events from muon neutrinos of that GRB falls below 1event/$km^{2}$. 
When the order of the muon threshold energy is PeV the number muon events from
 individual GRBs does not change significantly due to the change in the value of minimum Lorentz factor.

\para
In Fig.3. the dependence of numbers of muon events from tau neutrinos and muon neutrinos of a GRB on the redshift of the GRB has been shown.The energy emitted in a burst is assumed to be $E_{GRB}=10^{54} erg$. The number of muon events falls rapidly as we 
increase the redshift of the GRB. The detector threshold 
energy is assumed to be $E_{th}=30$ PeV and the zenith angle of observation is
$\theta=180^{\circ}$. The GRBs are distributed uniformly upto a redshift of
about $z=2$ \cite{piran}. There is only opportunity of detecting tau neutrino
appearance from bursts at very low redshifts.
From Fig.3. we notice that at all redshifts the numbers of muon events from muon neutrinos and tau neutrinos of a GRB remain nearly equal for muon
threshold energy $E_{th}=30$PeV.

\para
The numbers of muon events from tau neutrinos and muon neutrinos of a GRB
 increase linearly with increasing values of $E_{GRB}$ which is the total energy emitted in neutrino emissions.  

We also find from our calculations that when $E_{GRB}=10^{53}erg$ and $E_{GRB}=10^{54}erg$ we can not expect to detect tau neutrino signals from individual GRBs in $km^{2}$ area muon detector if their redshifts are more than $z=0.01$ and $z=0.03$ respectively. 
We have used the allowed values of minimum Lorentz factor and photon spectral break energy for wind luminosity $10^{53}erg/sec$ from \cite{Guetta}. The wind durations are typically of the order of 10 seconds.
One can find similar constraints for detectability of tau neutrinos from individual GRBs in $km^{2}$ area muon detector for other values of $E_{GRB}$ using the allowed values of GRB parameters such as minimum Lorentz factor, photon spectral break energy and wind luminosity corresponding to those values of $E_{GRB}$.

\section{Conclusion}
The detectability of tau neutrinos produced from oscillations of muon neutrinos originated from individual GRBs has been tested for the first time in a muon
detector of $km^{2}$ area. The role of the GRB parameters on the detectability
of tau neutrinos from individual GRBs has been explained in this work.
If the energies emitted in neutrino emissions are $10^{53}erg$ and $10^{54}erg$ and the bursts are at distances of redshifts less than $z=0.01$ and $z=0.03$ respectively, it will be possible to detect $\nu_{\tau}$ signals from them in a $km^{2}$ area muon detector in the vertical direction. The muon threshold energy has to be 30 PeV for this purpose. Also the oscillation parameters influence the number of $\nu_{\tau}$ events. One can find out the values of neutrino mass difference and mixing parameters in extragalactic neutrinos by observing the disappearance of muon neutrinos or the appearance of tau neutrinos.

\section{Acknowledgment}
The author is thankful to the referee for helpful suggestions.

\begin{figure}
\newpage
\psfig{figure=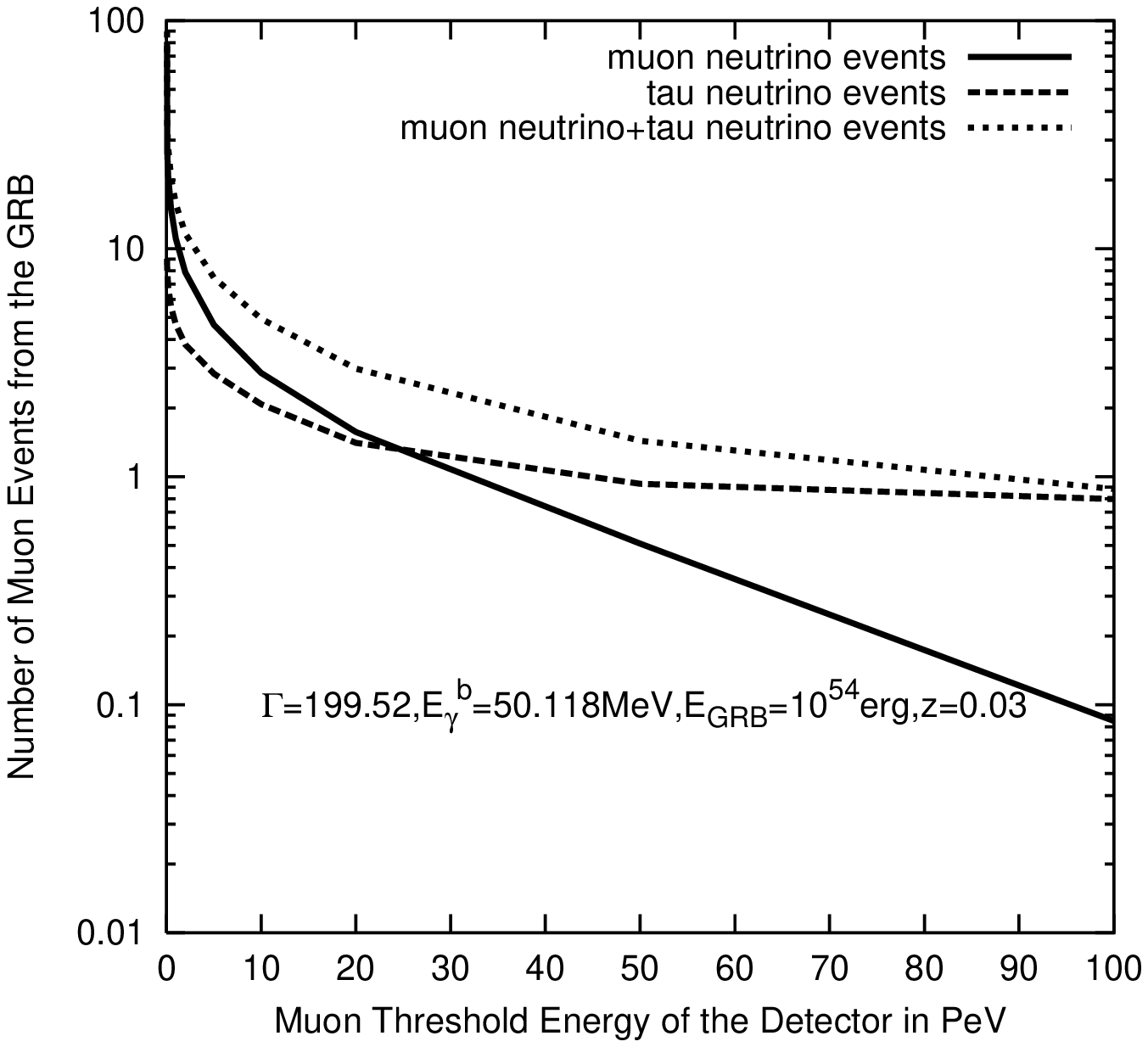, height=11cm} 
\caption{Number of secondary muon events produced from tau neutrinos of a GRB via tau lepton decays ($\nu_{\tau}\rightarrow {\tau}\rightarrow {\mu}$) in a muon detector of $km^2$ area and also the number of muon events from muon neutrinos of that GRB have been plotted in this figure for various threshold energies of the detector. The GRB is at a zenith angle of $\theta=180^{\circ}$. The wind luminosity of the GRB is assumed to be $10^{53}erg/sec$ for equipartition parameter $\epsilon_{B}=0.01$. The correlation of GRB parameters $\Gamma$ and photon spectral break energy $E_{{\gamma},MeV}^{b}$ have been used from \cite{Guetta}. The total energy emitted in each burst is assumed to be $E_{GRB}=10^{54}erg$ and the redshift of the burst is $z=0.03$. The sums of the numbers of muon neutrino and tau neutrino events in the muon detector have also been shown in this figure for different muon threshold energies. }
\end{figure}

\begin{figure}
\psfig{figure=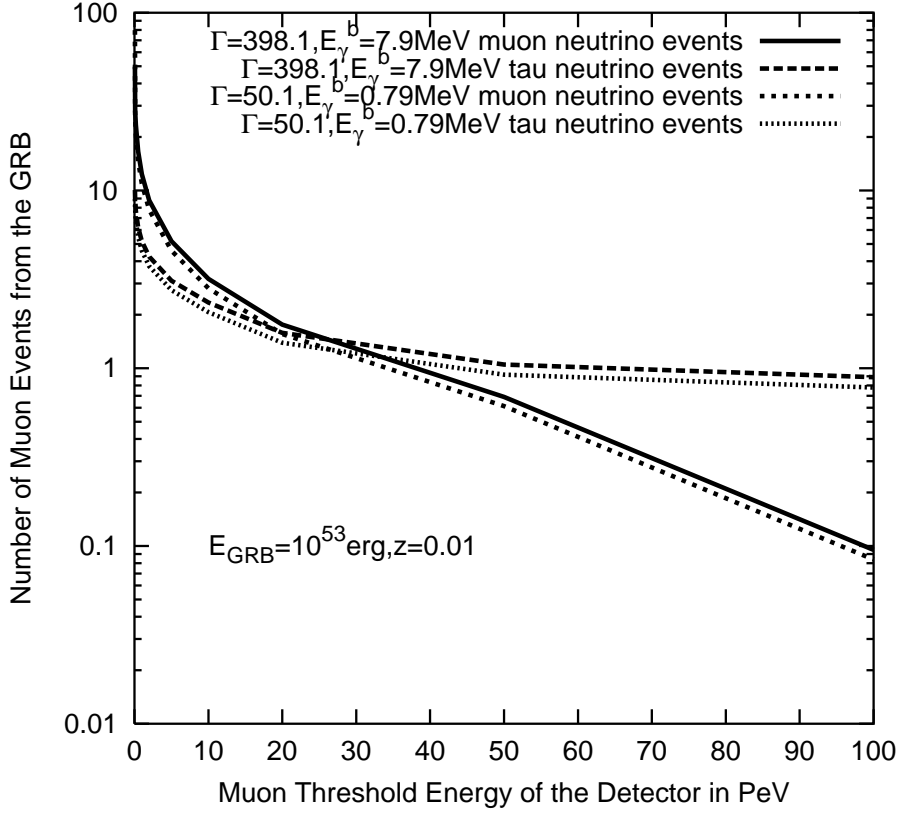,height=11cm} 
\caption{The numbers of muon events from $\nu_{\tau}$s and $\nu_{\mu}$s of a GRB in $km^{2}$ area have been plotted as a function of muon threshold energy of the detector. We have not considered the double bang events in our calculations.
 The burst is at $\theta=180^{\circ}$ and wind luminosity of the GRB
is assumed to be $10^{53}erg/sec$ for equipartition parameter $\epsilon_{B}=0.01$. The total energy emitted in neutrinos is assumed to be $E_{GRB}=10^{53}erg$. Here  $\Gamma$ is the Lorentz factor, and $E_{{\gamma},MeV}^{b}$ is the observed photon spectral break energy. Their allowed values are taken from \cite{Guetta}.}
\end{figure}

\begin{figure}
\psfig{figure=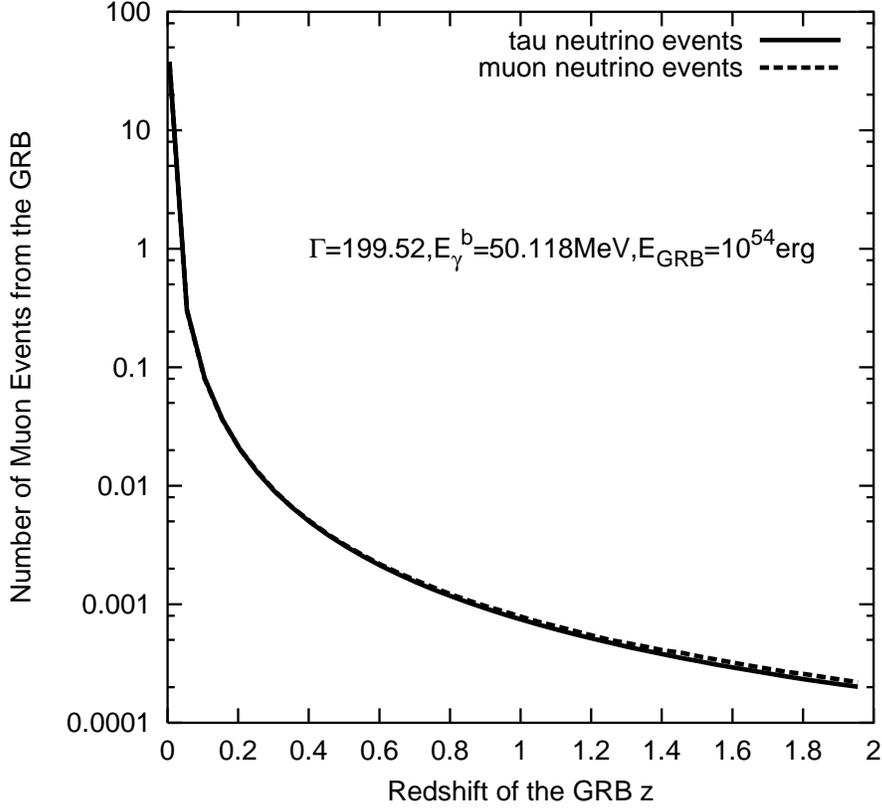,height=11cm}
\caption{The dependence of numbers of muon events produced via tau lepton decays ($\nu_{\tau}\rightarrow {\tau}\rightarrow {\mu}$) from tau neutrinos of a GRB  and also from muon neutrinos of that GRB on the redshift of the GRB has been shown  in this figure. The area of the muon detector is assumed to be $1 km^{2}$. The GRB is  at zenith angle $\theta=180^{\circ}$. The wind luminosity of GRB is assumed to be $10^{53}erg/sec$ and equipartition parameter $\epsilon_{B}=0.01$. The threshold energy of the detector is assumed to be $E_{th}=30$PeV. }
\end{figure}
\end{document}